%% file: main.tex
\documentclass[11pt]{article}

\usepackage[final]{acl}

\usepackage{times}
\usepackage{latexsym}

\usepackage[T1]{fontenc}

\usepackage[utf8]{inputenc}

\usepackage{microtype}

\usepackage{inconsolata}

\usepackage{graphicx}
\usepackage{enumitem}
\usepackage{booktabs}
\usepackage{array}
\usepackage{tabularx}
\usepackage{pifont}
\usepackage{makecell}
\usepackage{stfloats}
\usepackage{arydshln}
\usepackage{multirow}
\usepackage{float}

\title{SpeechEditBench: A Bilingual Multi-Attribute Benchmark for Instruction-Guided Speech Editing}

\author{
    \textbf{Hanlin ZHANG}\textsuperscript{1,3,*}, 
    \textbf{Daxin Tan}\textsuperscript{2,*}, 
    \textbf{Dehua Tao}\textsuperscript{2}, 
    \textbf{Xiao Chen}\textsuperscript{2, $\dagger$}, \\ 
    \textbf{Haochen Tan}\textsuperscript{2}, 
    \textbf{Linqi Song}\textsuperscript{1,3,$\dagger$} \\
    \textsuperscript{1}Department of Computer Science, City University of Hong Kong \\
    \textsuperscript{2}AI Lab, Leibniz Research Center, Huawei \\
    \textsuperscript{3}City University of Hong Kong Shenzhen Research Institute \\
    \texttt{hanlzhang8-c@my.cityu.edu.hk}, \\
    \texttt{\{tan.daxin1, TaoDehua, chen.xiao2, tan.haochen2\}@huawei.com} \\
    \texttt{linqi.song@cityu.edu.hk}, \\
    {\small \textsuperscript{*}Equal contribution. \textsuperscript{$\dagger$}Corresponding authors.}
}

\begin{document}
\maketitle

\begin{abstract}
Instruction-guided speech editing requires a model to modify specified speech attributes while preserving non-target characteristics. Despite rapid progress in Speech Large Language Models (Speech LLMs), systematic evaluation of this capability remains challenging, as existing benchmarks are fragmented across isolated editing tasks. To bridge this gap, we introduce \textbf{SpeechEditBench}, a bilingual multi-attribute benchmark for instruction-guided speech editing. SpeechEditBench encompasses seven atomic editing tasks, as well as compositional editing tasks that integrate multiple operations within a single instruction. We propose an anchor-based evaluation protocol that separately assesses the edit success of target attributes and the preservation of non-target linguistic content, leading to three metrics: target success, preservation success, and joint success. Using this benchmark, we evaluate mainstream Speech LLMs and specialized speech editing systems. The results reveal three key findings: (1) no single model performs well across all editing dimensions; (2) closed-source Speech LLMs generally outperform open-source models; (3) compositional editing poses a significant challenge, with even the most advanced models struggling to achieve high joint success. SpeechEditBench provides a rigorous diagnostic framework to identify bottlenecks in Speech LLMs, thereby facilitating the development of next-generation models with more precise instruction-guided editing capabilities. Data and code are available at \url{https://github.com/daxintan-cuhk/SpeechEditBench}.

\end{abstract}

\section{Introduction}
Speech large language models (Speech LLMs) have rapidly become the dominant paradigm for modern speech intelligence, delivering strong performance across diverse tasks under a unified model architecture~\cite{ding2025kimi,zhang2025mimo,wu2025step,chu2024qwen2,arora2025landscape}. Current Speech LLMs have achieved remarkable progress in mainstream applications, including automatic speech recognition, text-to-speech synthesis, speech translation, and spoken question answering~\citep{llmbasedasr,llasa,ding2025kimi}. These well-established tasks are supported by abundant public datasets, mature evaluation benchmarks, and standardized testing protocols, enabling fair model comparison and steady technical advancement.

In contrast, instruction-guided speech editing remains an under-explored capability in the Speech LLM research landscape. Unlike conventional speech tasks with straightforward and standalone objectives, speech editing imposes dual constraints: it demands targeted attribute modification while preserving unrelated speech characteristics. The model is required to comprehend natural language editing instructions, locate target attributes in source utterances, perform precise semantic and acoustic manipulation, and faithfully preserve all unchanged content and speaker characteristics. Speech editing also serves as a comprehensive diagnostic task, evaluating whether modern Speech LLMs can consolidate isolated editing capabilities originally scattered across multiple domain-specific expert systems.

Unlike recognition~\cite{librispeech}, synthesis~\cite{zen2019libritts}, and question answering~\citep{audioqa}, which have well-established benchmarks, Speech-LLM-based speech editing lacks a dedicated, unified evaluation benchmark. Existing audio-editing benchmarks~\cite{yan2025step,yan2025ming}, however, use different task definitions and evaluation metrics, making comprehensive evaluation and systematic diagnosis of model strengths and weaknesses difficult. This fragmentation hinders progress in the field.
Furthermore, existing evaluation paradigms have several limitations. First, task-specific evaluation metrics are not directly comparable across studies. Second, rigid waveform matching cannot accommodate the one-to-many nature of valid speech edits. Third, existing evaluations rarely consider both editing effectiveness and source preservation simultaneously. To bridge this gap, our benchmark introduces a unified, fine-grained evaluation framework for instruction-guided speech editing.

\begin{itemize}[leftmargin=*, nolistsep]
\item We construct SpeechEditBench, to our knowledge the first \textbf{bilingual multi-attribute benchmark} for instruction-guided speech editing. It covers \textbf{seven atomic editing tasks} (content, speaker, emotion, style, prosody, paralinguistic, acoustic) and \textbf{compositional editing} (combining multiple operations).
\item We propose an \textbf{anchor-based evaluation protocol} that avoids rigid waveform matching, along with metrics that separately measure edit effectiveness (target success) and the preservation of non-target linguistic content (preservation success), as well as their conjunction (joint success).
\item Systematic evaluation of eight Speech LLMs and specialized systems reveals three main findings. First, no single model performs well across all editing dimensions. Second, closed-source Speech LLMs outperform open-source models on most tasks. Third, compositional editing remains highly challenging: even the strongest models achieve very low joint success. Additional analyses further reveal model-dependent language bias.
\end{itemize}

\section{Related Work}

\subsection{Instruction-Guided Editing Benchmarks} 
Instruction-guided editing requires balancing precise instruction execution with preservation of original content. This dual-objective challenge has been extensively studied in the image domain. Foundational works include EditBench~\cite{wang2023imagen} which evaluates text-guided image inpainting across objects, attributes, and scenes, and EditVal~\cite{basu2023editval} a standardized benchmark with 13 edit types and automated VLM evaluation. I\textsuperscript{2}EBench~\cite{ma2024i2ebench} introduced a 16-dimension evaluation framework covering high-level and low-level editing. Complex-Edit~\cite{yang2025texttt} enables complexity-controllable evaluation via chain-of-edit instruction generation, and EditInspector~\cite{yosef2025editinspector} provides fine-grained verification of accuracy, artifact detection, visual quality, and scene integration. Beyond text guidance, emerging benchmarks explore multimodal interaction: GIE-Bench~\cite{qian2025gie} evaluates functional correctness and content preservation via object-aware masking; VIBE~\cite{zhang2026well} assesses visual instruction-driven editing with sketches and manipulative vectors; SpotEdit~\cite{ghazanfari2025spotedit} focuses on hallucination in visually-guided editing. 

\subsection{General Speech and Audio Benchmarks}
Beyond traditional speech recognition and audio captioning, Speech LLM benchmarks have evolved to probe complex interaction and reasoning. VoiceBench~\cite{chen2026voicebench} evaluates robustness to speaker characteristics, noise, and disfluencies. AIR-Bench~\cite{yang2024air} evaluates generative comprehension across speech, natural sounds, and music, whereas AudioBench~\cite{wang2025audiobench} targets speech, audio-scene, and voice or paralinguistic understanding. VocalBench~\cite{liu2025vocalbench} assesses conversational semantic quality, acoustic performance, conversational abilities, and robustness. For high-level reasoning, MMSU~\cite{wang2025mmsu} covers 47 linguistically grounded tasks across phonetics, prosody, syntax, and semantics, while MMAR~\cite{ma2026mmar} requires multi-step reasoning in mixed-modality scenarios and provides chain-of-thought rationales. For real-time interaction, Full-Duplex-Bench~\cite{lin2025full} evaluates turn-taking including pause handling, backchanneling, and interruption management; MTR-DuplexBench~\cite{zhang2025mtr} extends to multi-round evaluation of dialogue quality and dynamics. Collectively, these benchmarks measure how models comprehend and interact with speech. In contrast, our work focuses on instruction-driven speech editing across speaker, content, acoustic, and prosodic attributes.

\subsection{Speech Editing Systems and Evaluation}
Early text-based speech editing (TSE) systems such as EditSpeech~\cite{tan2021editspeech} and FluentEditor~\cite{liu2023fluenteditor,liu2024fluenteditor2} primarily framed the task as spectrogram inpainting, focusing on boundary smoothness and local prosody. Neural codec language models later enabled end-to-end token-infilling editors such as VoiceCraft~\cite{peng2024voicecraft}, which introduced the RealEdit dataset, and VoiceCraft-X~\cite{zheng2025voicecraft}, which extends multilingual editing to 11 languages. CosyEdit~\cite{chen2026cosyedit} is a 400M-parameter model fine-tuned from CosyVoice, while MAVE~\cite{mohammad2025speak} adopts a cross-attentive Mamba architecture. Beyond these task-specific models, recent Speech LLMs have begun unifying editing with broader understanding and generation. Ming-UniAudio~\cite{yan2025ming} enables free-form instruction-guided editing via a continuous tokenizer, while Step-Audio-EditX~\cite{yan2025step} supports expressive iterative control over emotions and speaking styles within a 3B-parameter LLM. Despite these advances, evaluation remains fragmented, as surveyed in a comprehensive review~\cite{kassmann2024speech}. ISSE~\cite{chen2026isse} provides a large instruction-guided style-editing dataset and benchmark, while LibriSpeech-Edit~\cite{lv2026ast} introduces fine-grained evaluation based on word-level alignment.
Two instruction-guided audio-editing benchmarks, Ming-Freeform-Audio-Edit~\cite{yan2025ming} and Step-Audio-Edit-Benchmark~\cite{yan2025step}, cover partially overlapping attribute families but use different task taxonomies and do not evaluate the same bilingual compositional setting as SpeechEditBench.
Moreover, precise multi-attribute editing---simultaneously modifying speaker identity, linguistic content, acoustic characteristics, and prosody---remains an open challenge.

\section{Benchmark Design}

\begin{figure*}[!htb]
\centering
\includegraphics[trim=90 100 80 90, clip, width=\textwidth]{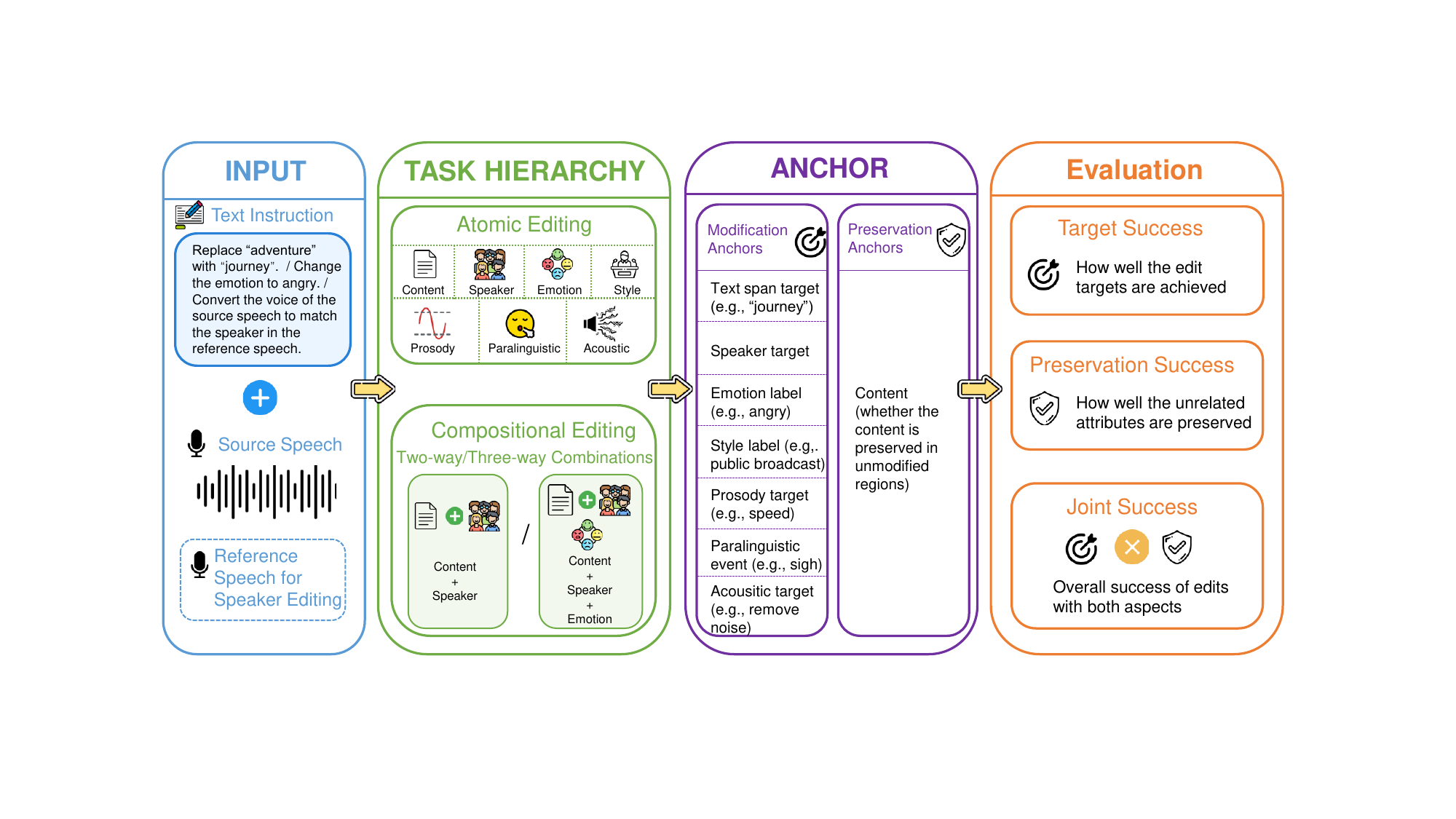}
\caption{Overall framework of the proposed SpeechEditBench.}
\label{fig:framework}
\end{figure*}

\textbf{SpeechEditBench} is built on three principles: unified task formulation across all editing dimensions, anchor-based evaluation to avoid rigid waveform matching, and dual-constraint metrics that balance editing effectiveness and the preservation of linguistic content. Figure~\ref{fig:framework} illustrates the overall framework.

\subsection{Input Formulation and Task Hierarchy}
Each sample consists of a source speech and a natural language instruction, while for speaker editing an additional reference speech is provided as the target timbre. Models are requested to modify only the instructed attributes while preserving non-target linguistic content. The benchmark covers English and Chinese to evaluate cross-lingual generalization.
Two difficulty levels are defined: \textbf{Atomic editing} focuses on a single attribute per instruction to assess fundamental editing capability. \textbf{Compositional editing} combines two or three editing operations in one instruction to test multi-constraint satisfaction and joint instruction fulfillment.

\subsubsection{Atomic Editing Tasks}
We define seven atomic editing tasks, whose editing targets, anchor types, and main evaluation metrics are summarized in Table~\ref{tab:task_overview}.

\paragraph{Content Editing.}
Insertion, replacement, or deletion of words or phrases in the utterance.

\paragraph{Speaker Editing.}
Voice conversion that changes the timbre of the source speech to match a reference speaker.

\paragraph{Emotion Editing.}
Transformation of the source emotion into a target emotion. Two subsets with different text polarity are designed. In the \textit{standard set}, the source text is emotionally neutral. In the \textit{challenging set}, the source text expresses the same emotion in source speech (e.g., the words ``I am so angry'' with angry speech), but the instruction asks for a different target emotion (e.g., happiness). In this case, the model must override the semantic polarity of the text to realize emotion editing.

\paragraph{Style Editing.}
Transformation of the utterance into a target speaking style. While there is no universally accepted taxonomy for speaking styles, we adopt a pragmatic set of six distinct styles to cover a broad range of expressive scenarios: public-broadcast, intimate, dramatic, restrained-flat, storytelling, and conversational.

\paragraph{Prosody Editing.}
Adjustment of speed, pitch, or stress on specific words.

\paragraph{Paralinguistic Editing.}
Insertion of non-verbal vocal events (breath, laugh, cough, sigh) into a clean utterance or removal of such events from a given speech while keeping the main content intact.

\paragraph{Acoustic Editing.}
This task comprises two subtasks. \textit{Speech enhancement} removes background noise or dereverberates the audio. \textit{Environment transfer} adds background sounds (crowd, outdoor, music) or reverberation (bedroom, small room, hall) to a clean speech signal.

\begin{table*}[!htb]
\centering
\caption{Seven atomic editing tasks with editing targets, anchor types, and main metrics.}
\label{tab:task_overview}
\resizebox{\linewidth}{!}{
\begin{tabular}{llll}
\toprule
Task & Editing Target & Anchor Type & Main Metrics \\
\midrule
Content Editing & Replacement, insertion, deletion & Text span & WER/CER \\
Speaker Editing & Voice conversion & Speaker reference & Speaker similarity \\
Emotion Editing & Emotion conversion & Emotion label & Emotion classification accuracy \\
Style Editing & Speaking style conversion & Style label & Style classification accuracy \\
Prosody Editing & Speed, pitch and stress adjustment & Prosodic range & duration ratio, F0 shift, prominence gain \\
Paralinguistic Editing & Add or remove non-verbal events & Event category & Event detection accuracy \\
Acoustic Editing & Speech enhancement or environment transfer & Acoustic condition & DNSMOS gain, RT60, acoustic scene match \\
\bottomrule
\end{tabular}
}
\end{table*}

\subsubsection{Compositional Editing}
We group atomic edits into four categories: \textit{semantic content} (content editing), \textit{speaker identity} (speaker editing), \textit{expressive delivery} (emotion, style, prosody, paralinguistic), and \textit{acoustic environment} (acoustic editing). Based on these, we create a compositional split with 320 two-component and 80 three-component cross-category samples, balanced across English and Chinese. This design enables the evaluation of per-component success and joint instruction fulfillment.

\subsection{Dataset Construction}
\begin{figure}[!htb]
\centering
\includegraphics[trim=100 10 100 15, clip, width=\columnwidth]{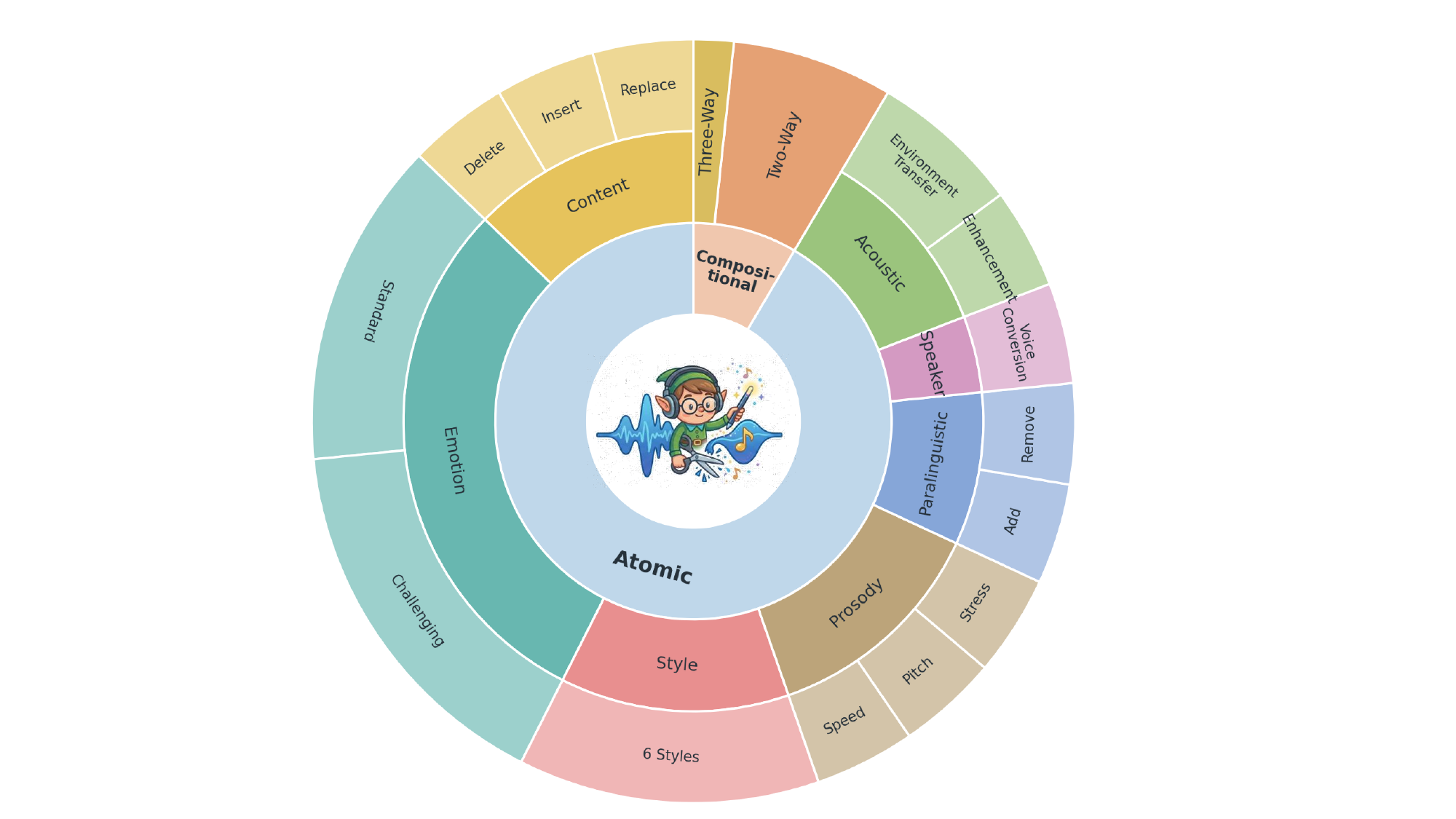}
\caption{Hierarchical sample distribution of \textbf{SpeechEditBench} (sunburst chart).}
\label{fig:sunburst}
\end{figure}

\textbf{SpeechEditBench} comprises 4,700 source--instruction pairs, with its hierarchical composition shown in Figure~\ref{fig:sunburst}. Audio is sourced from public English and Chinese corpora, including LibriTTS~\cite{zen2019libritts}, AISHELL-3~\cite{shi2020aishell}, WenetSpeech~\cite{zhang2022wenetspeech}, VCTK~\cite{yamagishi2019cstr}, IEMOCAP~\cite{busso2008iemocap}, CSEMOTIONS~\cite{tian2025marco}, NonverbalTTS~\cite{borisov2025nonverbaltts}, DisfluencySpeech~\cite{wang2024disfluencyspeechsinglespeakerconversational}, LibriQuote~\cite{michel2025libriquote}, NaturalVoices~\cite{du2025naturalvoices}, Aishell6-whisper~\cite{li2026aishell6}, MagicData-RAMC~\cite{yang2022open}, StoryTTS~\cite{liu2024storytts}, and Emilia-NV~\cite{liao2026emilia}. MUSAN noises~\cite{snyder2015musan} and RIRS\_NOISES~\cite{ko2017study} are used to synthesize degraded speech with noise or reverberation.

Each sample has task-specific anchors (e.g., target transcripts, reference speakers, emotion/style labels, prosody directions, paralinguistic events, acoustic conditions). GPT-4o~\cite{openai2024gpt4o} is used for semantic edit proposal, text-based filtering, and free-form instruction generation. Gemini 2.5 Pro~\cite{google2026geminipro} assists with audio-based annotation and candidate selection for style and paralinguistic editing. For unambiguous targets, template-based instructions are used to keep prompts aligned with the annotated edit goal.

\subsection{Evaluation Protocol}

We adopt anchor-based evaluation. For each sample, target success $t_i\in\{0,1\}$ indicates whether the requested edit is correctly applied; preservation success $p_i\in\{0,1\}$ indicates whether the expected linguistic content is retained. Joint success is $j_i = t_i \cdot p_i$. Atomic scores are averaged over valid evaluator records and reported as percentages; missing component results in compositional editing are treated as failures.

For atomic content editing, joint success equals target success. For non-content atomic tasks and compositional tasks without content editing, the expected transcript is the source transcript; for compositional tasks with content editing, it is the corresponding target transcript. We transcribe the model output with ASR and set $p_i=1$ if its WER/CER against the expected transcript does not exceed $10\%$.

Target success criteria are task-adaptive with predefined thresholds. \textbf{Content editing} requires the target span to be present (or absent for deletion) via ASR alignment. \textbf{Speaker editing} requires cosine similarity $\ge 0.50$ between output and reference speaker embeddings. \textbf{Emotion, style, and paralinguistic editing} use a Gemini-based judge: emotion requires label match; style requires target style score $\ge 3$; paralinguistic requires event score $\ge 2$ for addition or $\le 1$ for removal. \textbf{Prosody editing} uses duration ratio ($\le 0.95$ for faster, $\ge 1.05$ for slower), median F0 shift ($\ge +0.3$ or $\le -0.3$ semitone), and prominence gain for stress. \textbf{Acoustic editing} requires positive DNSMOS gains for enhancement, RT60 within 0.8--1.2 times target for reverb transfer, or PANNs scene prediction matching the target with score gain $\ge 0.03$ for noise transfer.

For compositional samples, we evaluate each component with the corresponding atomic metric and report component-wise success (proportion of components where target success is achieved), all-component success (all components in a sample succeed), and joint success (all-component success with preservation constraints applied). Full metric definitions, judge prompts, and thresholds are provided in Appendix~\ref{app:evaluation-details}.

\section{Evaluated Models}

We evaluate two distinct model categories for comprehensive benchmark assessment.

\textbf{Speech LLMs.}
This is our primary target for general cross-task editing capability. We evaluate six open-source systems: Ming-UniAudio~\cite{yan2025ming}, Step-Audio-EditX~\cite{yan2025step}, Qwen3-Omni~\cite{xu2025qwen3}, Kimi-Audio~\cite{ding2025kimi}, Mimo-Audio-Base and Mimo-Audio-Instruction~\cite{zhang2025mimo}. We also evaluate two closed-source systems, Gemini-Live~\cite{google2026geminilive} and GPT-Realtime~\cite{openai2026realtime}. Speaker editing is not evaluated because the model interfaces used in this study do not provide a standardized source-plus-reference editing workflow.

\textbf{Specialized speech editing systems.}
We also include task-specific systems as reference systems when their native interfaces align with a SpeechEditBench task. Specifically, we use VoiceCraft-X~\cite{zheng2025voicecraft} for content editing, Seed-VC~\cite{liu2024zero} for speaker editing, and VoxCPM2~\cite{voxcpm2_2026} for emotion, style, and prosody editing. For paralinguistic editing, we use Chatterbox.\footnote{
\url{https://github.com/resemble-ai/chatterbox}} for event addition and AudioSep~\cite{liu2023separate} for event removal. For acoustic editing, we use DeepFilterNet~\cite{schroeter2023deepfilternet3} for speech enhancement, while procedural noise mixing and RIR convolution implement environment transfer. These systems are not unified instruction-following editors but provide task-specific reference performance.

\begin{table*}[!htb]
\centering
\scriptsize
\setlength{\tabcolsep}{3.5pt}
\begin{tabular}{lcccccccc}
\toprule
Model & Content $\uparrow$ & Speaker $\uparrow$ & Emotion $\uparrow$ & Style $\uparrow$ & Prosody $\uparrow$ & Paralinguistic $\uparrow$ & Acoustic $\uparrow$ & Compositional $\uparrow$ \\
\midrule
\multicolumn{9}{l}{\textbf{Open-source Speech LLMs}} \\
\addlinespace[0.2em]
Ming-UniAudio & 76.46 & N/T & 3.43 (5.29) & 22.17 (32.50) & 26.50 (28.00) & 11.25 (29.25) & 25.85 (29.66) & 1.75 (14.77) \\
Step-Audio-EditX & 16.50 & N/T & 7.71 (9.29) & 49.67 (54.00) & 20.13 (51.51) & 31.25 (61.75) & 22.89 (40.96) & 2.00 (16.14) \\
Qwen3-Omni & 72.00 & N/T & 1.64 (15.29) & 24.17 (63.50) & 38.17 (44.83) & 14.50 (44.75) & \textbf{37.80} (42.00) & 5.00 (31.59) \\
Kimi-Audio & 34.67 & N/T & 2.36 (22.43) & 24.50 (63.33) & 13.50 (40.33) & 9.25 (55.00) & 8.25 (38.23) & 1.50 (16.02) \\
Mimo-Audio-Base & 31.67 & N/T & 0.21 (8.79) & 5.83 (42.17) & 5.67 (36.17) & 1.00 (49.75) & 4.44 (43.35) & 1.75 (15.00) \\
Mimo-Audio-Instruction & 64.17 & N/T & 0.86 (\textbf{42.36}) & 0.50 (77.50) & 0.67 (42.83) & 2.00 (67.50) & 0.80 (\textbf{45.69}) & 7.25 (32.05) \\
\midrule
\multicolumn{9}{l}{\textbf{Closed-source Speech LLMs}} \\
\addlinespace[0.2em]
Gemini-Live & 93.17 & N/T & \textbf{27.79} (34.43) & 63.67 (\textbf{84.00}) & \textbf{65.17} (69.67) & 26.50 (61.75) & 36.69 (41.53) & \textbf{11.00} (\textbf{38.52}) \\
GPT-Realtime & \textbf{96.67} & N/T & 14.57 (21.00) & \textbf{68.67} (82.33) & 63.94 (\textbf{70.12}) & \textbf{47.00} (\textbf{81.50}) & 27.60 (43.60) & 10.00 (34.89) \\
\midrule
\textbf{Specialized models} & \shortstack{EN 84.00\\ZH 47.00} & 80.50 (86.00) & 3.57 (4.14) & 32.67 (36.67) & 49.00 (50.83) & 19.25 (52.75) & 68.20 (73.20) & N/T \\
\bottomrule
\end{tabular}
\caption{\textbf{SpeechEditBench} main results. All scores are percentages. Each cell reports joint success, with target success in parentheses; for compositional editing, the parenthesized value is component success on the full 400-sample split. Content editing has identical joint and target success, so only one score is shown. Bold marks the best Speech LLM score in each column, separately for the primary and parenthesized scores. N/T indicates not tested; Speech LLM speaker editing is not evaluated because the interfaces used in this study do not provide a standardized source-plus-reference editing workflow. VoiceCraft-X content results are reported separately for English and Chinese due to a large language gap.}
\label{tab:main-results}
\end{table*}

\begin{figure*}[!htb]
\centering
\includegraphics[width=\textwidth]{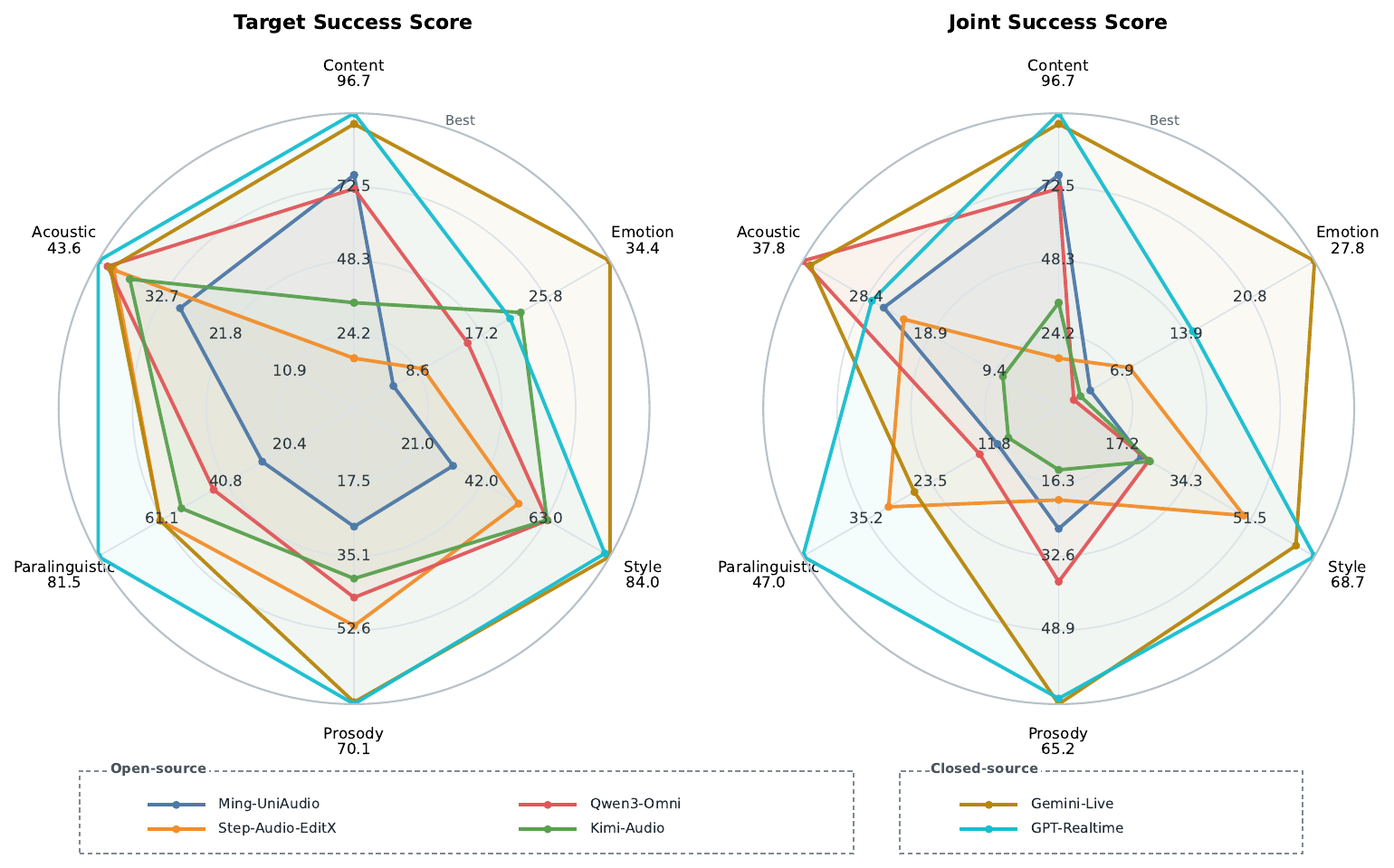}
\caption{Task-wise capability profiles of six representative Speech LLMs, including four open-source models and two closed-source models. The radar charts cover six atomic editing tasks, excluding speaker editing; values on each task axis are normalized by the highest Speech LLM score for that task. Target success measures whether the requested edit is achieved, while joint success additionally requires linguistic-content preservation.}
\label{fig:radar-results}
\end{figure*}

\section{Results and Discussion}

Table~\ref{tab:main-results} presents the main results across all evaluated models. Each cell reports joint success with target success (or component success for compositional editing) in parentheses; for content editing, joint and target success are identical, so only one score is shown. This format exposes the target--preservation gap: several models achieve the requested edit but fail to preserve the expected linguistic content. Speaker editing is marked as N/T for Speech LLMs because the evaluated interfaces do not provide a standardized source-plus-reference editing workflow. Figure~\ref{fig:radar-results} visualizes task-wise capability profiles for six representative Speech LLMs. The target-success radar shows that several models can perform the requested edits across multiple tasks, while the joint-success radar is much more contracted, revealing linguistic-content preservation as a central bottleneck.

\subsection{Human Validation of Judge-Dependent Metrics}
\label{sec:human-validation}

We validate the Gemini-dependent emotion, style, and paralinguistic metrics on a stratified subset of 240 outputs. The subset covers both languages, target-success strata, task subtypes, and nine systems for which outputs were available. Ten annotators each evaluated 72 anonymized items, yielding three independent annotations per output and 720 item-level annotations in total. For each output, annotators judged (i) whether the target edit succeeded and (ii) whether linguistic content was preserved; human joint success is their conjunction. Automatic decisions are compared with the majority label. Table~\ref{tab:human-validation} summarizes the results; Appendix~\ref{app:human-validation} gives the sampling, annotation, task-level statistics, and representative agreement and disagreement cases.

\begin{table}[H]
\centering
\small
\setlength{\tabcolsep}{5pt}
\begin{tabular}{lc}
\toprule
Validation metric & Result \\
\midrule
Human--human agreement & 80.3\% \\
Automatic--human agreement & 68.8\% \\
Cell-level Pearson $r$ & $0.685$ ($p<0.001$) \\
Cell-level Spearman $\rho$ & $0.699$ ($p<0.001$) \\
\bottomrule
\end{tabular}
\caption{Human validation of joint success on 240 outputs. Human--human is mean pairwise agreement; automatic--human agreement compares automatic decisions with majority labels. Correlations are computed over 27 task--system cells.}
\label{tab:human-validation}
\end{table}

For the primary joint criterion, automatic decisions agree with human majority labels on 68.8\% of outputs (Wilson 95\% CI: 62.6--74.3\%), with cell-level Pearson and Spearman correlations of $.685$ and $.699$ (both $p<.001$). These results demonstrate meaningful sample-level agreement and strong alignment in aggregate trends, supporting the use of the automatic metric for system-level joint comparisons.

\subsection{Open-source vs. Closed-source Speech LLMs}

Among open-source models, performance is highly fragmented. Ming-UniAudio leads content editing (76.46\% joint) but is weak on expressive tasks. Step-Audio-EditX achieves the best open-source results on emotion, style, and paralinguistic editing (7.71\%, 49.67\%, 31.25\% joint), though its joint success on content is low. Qwen3-Omni excels in prosody and acoustic editing (38.17\%, 37.80\% joint) while also performing strongly on content (72.00\%). The two Mimo variants show contrasting behavior: Mimo-Audio-Instruction reaches high target success on several tasks but often fails to preserve expected linguistic content, resulting in near-zero joint success; Mimo-Audio-Base generally has lower target success but achieves higher joint success on several tasks.

Closed-source systems substantially outperform open-source Speech LLMs on several dimensions, but they show different strengths. GPT-Realtime achieves the best content and style joint success among Speech LLMs (96.67\% and 68.67\%), and is also strongest on paralinguistic editing (47.00\%). Gemini-Live is strongest on emotion and prosody editing (27.79\% and 65.17\%), and remains highly competitive on content and style. On acoustic editing, Qwen3-Omni exceeds both closed-source systems in joint success, narrowly outperforming Gemini-Live (37.80\% vs. 36.69\%), showing that the closed-source advantage is not uniform across all tasks.

A key difference between open-source and closed-source systems is that the latter generally achieve higher scores on both target and joint success. For example, GPT-Realtime achieves 81.50\% target success and 47.00\% joint success on paralinguistic editing, while the strongest open-source model on this task, Step-Audio-EditX, reaches 61.75\% target success but only 31.25\% joint success. Similarly, Gemini-Live obtains 84.00\% target success and 63.67\% joint success on style editing, whereas Step-Audio-EditX obtains 54.00\% and 49.67\%, respectively. These results indicate that current open-source Speech LLMs still lag behind closed-source systems, especially in combining target control with preservation, although some remain competitive on individual tasks.

\subsection{Specialized Models vs. Speech LLMs}

In addition to Speech LLMs, we include task-specific specialized systems as reference systems. These systems are not unified editors, but they demonstrate the strength of mature task-specific pipelines. VoiceCraft-X is strong on English content editing (84.00\% joint) but much weaker on Chinese (47.00\%). Seed-VC achieves 80.50\% joint success on speaker editing under a reference-based setting, which is not evaluated for Speech LLMs because the interfaces used in this study lack a standardized source-plus-reference editing workflow. DeepFilterNet and the procedural acoustic editing methods achieve 68.20\% joint success overall, substantially outperforming the evaluated Speech LLMs on acoustic editing.

The comparison is more mixed for expressive tasks: VoxCPM2 is competitive on prosody (49.00\% joint) but weaker than the best Speech LLMs on emotion and style, while Chatterbox+AudioSep does not surpass the strongest models on paralinguistic editing. Thus, specialized systems remain strong references for several dimensions, especially speaker and acoustic editing, but they do not uniformly dominate Speech LLMs. Integrating their strengths into a unified instruction-following editor remains an open challenge.

\subsection{Target Success vs. Content Preservation}

\begin{table}[t]
\centering
\scriptsize
\setlength{\tabcolsep}{3.2pt}
\resizebox{\columnwidth}{!}{%
\begin{tabular}{lcccccc}
\toprule
\multirow{2}{*}{Task}
& \multicolumn{3}{c}{Ming-UniAudio}
& \multicolumn{3}{c}{Step-Audio-EditX} \\
\cmidrule(lr){2-4}
\cmidrule(lr){5-7}
& Joint & Target & Pres.
& Joint & Target & Pres. \\
\midrule
Emotion & 3.43 & 5.29 & 66.64 & 7.71 & 9.29 & 90.00 \\
Style & 22.17 & 32.50 & 66.33 & 49.67 & 54.00 & 88.67 \\
Prosody & 26.50 & 28.00 & 84.50 & 20.13 & 51.51 & 39.60 \\
Paralinguistic & 11.25 & 29.25 & 50.38 & 31.25 & 61.75 & 44.75 \\
Acoustic & 25.85 & 29.66 & 91.40 & 22.89 & 40.96 & 48.80 \\
\bottomrule
\end{tabular}%
}
\caption{Target--preservation trade-off for two representative open-source models on non-content, non-speaker atomic editing tasks. All scores are percentages. Pres. denotes content preservation.}
\label{tab:breakdown}
\end{table}

We focus on three metrics: target success, content preservation, and joint success. Table~\ref{tab:breakdown} compares two representative open-source models, Ming-UniAudio and Step-Audio-EditX, on non-content, non-speaker atomic tasks.

The two systems exhibit different failure modes. Step-Audio-EditX is stronger at changing expressive attributes, especially style and paralinguistic events, but its target advantage can be offset by poor preservation. On prosody and acoustic editing, it has higher target success than Ming-UniAudio but lower joint success because content preservation drops sharply. Ming-UniAudio shows the opposite tendency: it preserves content more reliably, especially on prosody and acoustic editing, but often changes the target attribute less successfully. This gap shows why target-only evaluation can overestimate speech editing ability.

\subsection{Compositional Editing Performance}

\begin{table}[t]
\centering
\footnotesize
\setlength{\tabcolsep}{2.2pt}
\renewcommand{\arraystretch}{1.0}
\begin{tabular}{lcccccc}
\toprule
\multirow{2}{*}{Model} & \multicolumn{3}{c}{2-component} & \multicolumn{3}{c}{3-component} \\
\cmidrule(lr){2-4}\cmidrule(lr){5-7}
& Comp. & All & Joint & Comp. & All & Joint \\
\midrule
Ming-UniAudio & 19.00 & 2.00 & 2.00 & 13.33 & 0.00 & 0.00 \\
Step-Audio-EditX & 22.00 & 4.00 & 3.50 & 23.33 & 0.00 & 0.00 \\
Qwen3-Omni & 40.50 & 10.50 & 10.00 & 43.33 & 0.00 & 0.00 \\
Kimi-Audio & 22.25 & 3.00 & 3.00 & 20.00 & 0.00 & 0.00 \\
Mimo-Audio-Base & 20.75 & 2.50 & 2.50 & 21.67 & 5.00 & 5.00 \\
Mimo-Audio-Inst. & 45.00 & 17.00 & 14.50 & 33.33 & 0.00 & 0.00 \\
Gemini-Live & 50.25 & 21.50 & 21.50 & 41.67 & 0.00 & 0.00 \\
GPT-Realtime & 49.75 & 20.50 & 20.00 & 40.00 & 0.00 & 0.00 \\
\bottomrule
\end{tabular}
\caption{Compositional editing on the Speech-LLM-compatible subset, excluding samples with speaker-editing components. All scores are percentages. Comp. is pooled per-component success; All requires all requested components in a sample to succeed; Joint additionally applies applicable preservation constraints.}
\label{tab:comp-results}
\end{table}

As the evaluated Speech LLMs do not accept reference audio, we exclude compositional samples containing speaker-editing components and analyze the remaining Speech-LLM-compatible subset. Table~\ref{tab:comp-results} shows that compositional editing remains difficult: component-level success is much higher than all-component or joint success, indicating that models often satisfy only part of a multi-intent instruction.

Closed-source models improve two-component compositional editing but do not solve it. Gemini-Live and GPT-Realtime reach around 50\% component success on two-component combinations, but their joint success remains only 21.50\% and 20.00\%, respectively. Open-source models are substantially lower, with Mimo-Audio-Instruction the strongest among them at 14.50\% joint success. 
Notably, nearly all models obtain 0.00\% joint success on this harder setting.

\begin{table}[t]
\centering
\footnotesize
\setlength{\tabcolsep}{3.0pt}
\resizebox{\columnwidth}{!}{%
\begin{tabular}{lcccccc}
\toprule
Model
& \multicolumn{3}{c}{\makecell{Content Editing Tasks\\Target Success}}
& \multicolumn{3}{c}{\makecell{Non-content Editing Tasks\\Content Preservation}} \\
\cmidrule(lr){2-4}
\cmidrule(lr){5-7}
& En & Zh & $\Delta$
& En & Zh & $\Delta$ \\
\midrule
Ming-UniAudio & 73.33 & 79.60 & +6.27 & 76.37 & 67.57 & -8.80 \\
Step-Audio-EditX & 1.67 & 31.33 & +29.67 & 69.03 & 55.67 & -13.36 \\
Qwen3-Omni & 73.00 & 71.00 & -2.00 & 65.55 & 59.77 & -5.78 \\
Kimi-Audio & 49.33 & 20.00 & -29.33 & 19.65 & 38.40 & +18.75 \\
Mimo-Audio-Base & 32.00 & 31.33 & -0.67 & 7.86 & 11.51 & +3.65 \\
Mimo-Audio-Inst. & 65.33 & 63.00 & -2.33 & 2.01 & 4.10 & +2.10 \\
\midrule
Gemini-Live & 98.33 & 88.00 & -10.33 & 82.38 & 72.05 & -10.33 \\
GPT-Realtime & 97.00 & 96.33 & -0.67 & 76.72 & 64.13 & -12.59 \\
\bottomrule
\end{tabular}%
}
\caption{Language effects on content editing (target success) and non-content editing (content preservation). $\Delta$ = Zh - En, in percentage points.}
\label{tab:language-bias}
\end{table}

\subsection{Language Bias Analysis}
\label{sec:language_bias}

Since \textbf{SpeechEditBench} is bilingual, we separate language effects according to the role of lexical content. In content editing, lexical content is the edit target; in non-content, non-speaker tasks, lexical content should be preserved while another speech attribute is modified. Table~\ref{tab:language-bias} reports target success for the former and content preservation for the latter (macro-averaged over emotion, style, prosody, paralinguistic, and acoustic editing).

The results show model-dependent language effects. For content editing, Ming-UniAudio and Step-Audio-EditX perform better on Mandarin, whereas Kimi-Audio and Gemini-Live perform better on English. Qwen3-Omni, Mimo-Audio-Base, Mimo-Audio-Instruction, and GPT-Realtime are comparatively balanced. For non-content tasks, content preservation often favors English: Ming-UniAudio, Step-Audio-EditX, Qwen3-Omni, Gemini-Live, and GPT-Realtime all show lower Mandarin preservation scores. Kimi-Audio and the two Mimo variants show the opposite trend, although their absolute preservation scores are much lower. These results show that aggregate bilingual scores can obscure whether language effects arise from executing lexical edits or preserving lexical content during non-content editing.

\subsection{Emotion Standard vs. Challenging Sets}
\label{sec:emotion_subset_analysis}

\begin{table}[t]
\centering
\footnotesize
\setlength{\tabcolsep}{2.8pt}
\resizebox{\columnwidth}{!}{%
\begin{tabular}{lcccccc}
\toprule
Model
& \multicolumn{3}{c}{Target Success Score}
& \multicolumn{3}{c}{Joint Success Score} \\
\cmidrule(lr){2-4}
\cmidrule(lr){5-7}
& Standard & Challenging & $\Delta$
& Standard & Challenging & $\Delta$ \\
\midrule
Ming-UniAudio & 6.62 & 4.13 & -2.49 & 5.08 & 2.00 & -3.08 \\
Step-Audio-EditX & 8.62 & 9.87 & +1.25 & 6.92 & 8.40 & +1.48 \\
Qwen3-Omni & 14.15 & 16.27 & +2.12 & 0.46 & 2.67 & +2.21 \\
Kimi-Audio & 29.08 & 16.67 & -12.41 & 3.69 & 1.20 & -2.49 \\
Mimo-Audio-Base & 9.54 & 8.13 & -1.41 & 0.15 & 0.27 & +0.12 \\
Mimo-Audio-Inst. & 45.23 & 39.87 & -5.36 & 0.46 & 1.20 & +0.74 \\
\midrule
Gemini-Live & 43.85 & 26.27 & -17.58 & 37.08 & 19.73 & -17.35 \\
GPT-Realtime & 22.77 & 19.47 & -3.30 & 14.31 & 14.80 & +0.49 \\
\bottomrule
\end{tabular}%
}
\caption{Emotion editing performance on standard (neutral text) and challenging (lexical-affective conflict) subsets. $\Delta$ denotes challenging - standard, in percentage points.}
\label{tab:emotion-standard-challenging}
\end{table}

The emotion split isolates text--speech modality conflict by comparing neutral-text examples with examples whose lexical affect conflicts with the target emotion. Table~\ref{tab:emotion-standard-challenging} shows that the challenging subset reduces average target success from 22.48\% to 17.59\%, suggesting that lexical-affective conflict generally makes emotion control harder.

The effect, however, is model-dependent. Kimi-Audio, Gemini-Live, and Mimo-Audio-Instruction show clear target-success drops, while Step-Audio-EditX and Qwen3-Omni slightly improve. Joint success highlights preservation as an additional bottleneck: Gemini-Live drops substantially on challenging examples, whereas GPT-Realtime remains stable, and open-source models remain low in both subsets. The split therefore helps distinguish general emotion-control difficulty from the additional difficulty associated with lexical-affective conflict.

\section{Conclusion}
We presented SpeechEditBench, a bilingual benchmark for instruction-guided speech editing, comprising 4,700 samples across seven atomic tasks and compositional editing. Its anchor-based evaluation protocol separately measures target success, preservation success, and joint success.

Evaluating eight Speech LLMs alongside task-specific systems reveals three main findings. First, current models exhibit severe capability fragmentation: no single model performs reliably across all editing dimensions. Second, linguistic-content preservation remains a central bottleneck: models that succeed at the requested edit often corrupt lexical content. Third, compositional editing is far from solved: even the strongest open-source model achieves low joint success. Human validation on 240 outputs further shows meaningful alignment between the automatic evaluation and human judgments on the primary joint criterion.

SpeechEditBench provides a unified testbed to diagnose these bottlenecks and support the development of Speech LLMs with more precise and reliable instruction-guided speech editing capabilities. The dataset and evaluation code are publicly available.

\section{Limitations}
\noindent\textbf{Coverage.} SpeechEditBench covers English and Chinese, seven atomic tasks, and their compositions, but excludes low-resource languages, code-switching, dialect conversion, voice-style interpolation, and multi-turn refinement.

\noindent\textbf{Evaluation.} The benchmark relies largely on automatic metrics. On 240 outputs, automatic joint decisions agree with majority human judgments on 68.8\%, while target-only agreement is lower. Human validation covers only the emotion, style, and paralinguistic judges and does not assess perceptual naturalness. Reliance on proprietary Gemini APIs for parts of annotation and evaluation may introduce evaluator dependence and affect reproducibility across API versions.

\noindent\textbf{Preservation.} The preservation gate measures only expected linguistic content, not other non-target characteristics such as speaker identity, prosody, or paralinguistic attributes; joint success therefore does not capture their preservation.

\section*{Acknowledgments}

This work was supported in part by the Research Grants Council of the Hong Kong SAR under GRF Grants 11217823 and 11216225 and Collaborative Research Fund Grant C1042-23GF; the National Natural Science Foundation of China under Grant 62371411; and the Laboratory for AI-Powered Financial Technologies under the InnoHK initiative of the Government of the HKSAR.


\bibliography{custom}

\appendix

\input{appendix}

\end{document}

%% file: appendix.tex
\appendix

\section{Construction and Evaluation Details}

This appendix provides supplementary methodological details that are not
expanded in the main paper: sample distributions, task construction criteria,
filtering prompts, evaluation thresholds, and qualitative failure types.

\subsection{Dataset Composition}

Table~\ref{tab:appendix-dataset-composition} summarizes the benchmark
composition. Counts are computed after filtering, balancing, and deduplication.
The benchmark contains 4,700 samples in total: 4,300 atomic samples and 400
compositional samples.

\begin{table*}[t]
\centering
\small
\setlength{\tabcolsep}{3.5pt}
\begin{tabular}{p{0.12\linewidth}p{0.06\linewidth}p{0.08\linewidth}p{0.30\linewidth}p{0.35\linewidth}}
\toprule
Task & Total & EN/ZH & Source datasets & Main distribution \\
\midrule
Content & 600 & 300/300 & LibriTTS test-clean 300; AISHELL-3 test 150; WenetSpeech test-net 150 & Replace 200, insert 200, delete 200; each language has 100 samples for each edit type. \\
Emotion & 1,400 & 750/650 & LibriTTS test-clean 350; AISHELL-3 test 300; IEMOCAP 400; CSEMOTIONS 350 & Standard 650, challenging 750. Target labels: angry 214, fearful 217, happy 218, sad 213, surprise 214, excited 109, frustrated 107, playfulness 108. \\
Style & 600 & 300/300 & LibriQuote 153; LibriTTS test-clean 43; NaturalVoices-VC 104; AISHELL-3 56; AISHELL6-Whisper 50; MagicData-RAMC 45; StoryTTS 90; WenetSpeech test-net 59 & Six source styles and six target styles are balanced at 100 samples each. Within each language, every source$\rightarrow$target direction has 10 samples. \\
Prosody & 600 & 300/300 & LibriTTS test-clean 300; AISHELL-3 test 300 & Speed 200, pitch 200, stress 200. Direction counts are faster 100, slower 100, higher 100, lower 100, stress 200. \\
Paralinguistic & 400 & 200/200 & LibriTTS test-clean 100; DisfluencySpeech 61; NonverbalTTS 39; AISHELL-3 test 50; WenetSpeech test-net 50; Emilia non-verbal ZH 100 & Add 200 and remove 200. Events are breath 100, laugh 100, cough 100, sigh 100. Each language--operation--event cell has 25 samples. \\
Speaker & 200 & 100/100 & VCTK 100; AISHELL-3 test 100 & Source and target speakers are always different. Each sample has a target-speaker reference clip of roughly 3--8 seconds. \\
Acoustic & 500 & 250/250 & LibriTTS test-clean 250; AISHELL-3 test 250, with MUSAN noises and synthetic room impulse responses for acoustic simulation & Enhancement 200, including noisy 100 and reverberant 100. Environment transfer 300, including reverb 150 and noise 150; each environment subtype has 50 samples. \\
Compositional & 400 & 200/200 & Derived from atomic-task sources: LibriTTS test-clean 140; AISHELL-3 test 135; WenetSpeech test-net 55; VCTK 50; IEMOCAP 10; CSEMOTIONS 10 & Two-component 320 and three-component 80. Each two-component combination has 40 samples; each three-component combination has 20 samples. \\
\midrule
Total & 4,700 & \makecell{2,400/\\2,300} & -- & -- \\
\bottomrule
\end{tabular}
\caption{Sample composition of SpeechEditBench. Counts refer to benchmark samples, not raw corpus sizes.}
\label{tab:appendix-dataset-composition}
\end{table*}

The source data can be grouped into five categories. Clean read speech supports
content, prosody, acoustic, and neutral expressive editing. Emotional and
expressive corpora provide emotion- and style-rich utterances. Speaker-labeled
multi-speaker corpora provide identity anchors. Non-verbal speech corpora
provide paralinguistic events. Acoustic resources provide additive noises and
room responses used to synthesize controlled acoustic conditions.

\subsection{Atomic Task Construction}

Table~\ref{tab:appendix-atomic-construction} gives the construction protocol
for each atomic task. In all cases, a sample contains a source speech signal, a
natural-language instruction, and task-specific anchors that define verifiable
targets.

\begin{table*}[t]
\centering
\small
\setlength{\tabcolsep}{3.5pt}
\begin{tabular}{p{0.12\linewidth}p{0.39\linewidth}p{0.39\linewidth}}
\toprule
Task & Source and instruction construction & Anchor construction and balancing \\
\midrule
Content & Source utterances are complete, self-contained transcripts of moderate length. A text-edit annotation identifies one replace, insert, or delete operation. Instructions are single-goal templates such as replacing a phrase, inserting a phrase after an anchor, or deleting a phrase. & Anchors record the original transcript, target transcript, edit type, edited span, and insertion anchor when applicable. The split is balanced by language and edit type. \\
Emotion & The standard subset uses neutral-text source utterances from clean read-speech corpora. The challenging subset uses utterances whose text itself is salient for a source emotion. Instructions specify only the target emotion. & Anchors record source emotion, target emotion, raw label, and taxonomy. Target emotions are assigned within the dataset taxonomy while avoiding invalid or same-label targets. \\
Style & Candidate utterances are annotated on six delivery-style dimensions by an audio judge. High-confidence utterances define the source-style examples. Instructions ask for conversion from the source style to a target style. & Anchors record source and target style. The six-style taxonomy is public-broadcast, intimate, dramatic, restrained-flat, storytelling, and conversational. Source styles, target styles, and source$\rightarrow$target directions are balanced. \\
Prosody & Speed and pitch examples use clean read speech. Stress examples additionally require a target word or phrase that can be localized in the transcript. Instructions are compact templates such as ``Speak faster'', ``Raise the pitch'', or ``Emphasize the word X''. & Anchors record prosody type and direction. Stress anchors record the target stress words. The split balances language and the three subtypes speed, pitch, and stress. \\
Paralinguistic & Add examples start from utterances where the target event is absent. Remove examples start from utterances where the target event is present. Instructions specify adding or removing one event. & Anchors record operation and event. Event labels are breath, laugh, cough, and sigh. The split balances language, operation, and event. \\
Speaker & Source and reference clips are selected from speaker-labeled corpora. Target speakers are assigned within the same language but are disjoint from the source speaker. Instructions refer to the reference clip rather than internal speaker IDs. & Anchors record source speaker, target speaker, and the target-speaker reference identity. Reference clips are constrained to a short usable duration, and pairings exclude same-speaker pairs. \\
Acoustic & Enhancement sources are synthesized degraded speech; environment-transfer sources are clean speech with a target acoustic environment. Instructions specify either removing noise/reverberation or adding a particular environment. & Anchors record enhancement degradation type or environment type/subtype. Reverb transfer records target RT60 ranges. Noise transfer records target environment subtype and SNR regime. \\
\bottomrule
\end{tabular}
\caption{Atomic task construction rules.}
\label{tab:appendix-atomic-construction}
\end{table*}

\subsection{Filtering Prompt Specifications}

Several tasks require filtering before sample construction. The following
prompt specifications summarize the criteria used by the annotation and
evaluation models. Language-specific variants use the same criteria and
structured response fields.

\paragraph{Content edit proposal.}
For each candidate utterance, the annotator first checks whether the text is
``a complete, self-contained utterance'' and rejects isolated names, titles,
pure numbers or codes, truncated sentences, and fragments lacking context. The
operation-specific prompt then asks for one feasible edit. For replacement, it
asks for one content word or short phrase of 1--3 words that can be naturally
replaced by a semantically related alternative while keeping the sentence
grammatical. For insertion, it asks for a natural insertion point and a
meaningful short phrase of 1--4 words. For deletion, it asks for one optional
word or short phrase that can be removed without breaking grammar or the core
meaning. The structured response contains the edited span, the target
transcript, a feasibility flag, a semantic-completeness flag, and a brief
reason.

\paragraph{Emotion filtering.}
The standard subset uses the neutral-text prompt:
\begin{quote}\small
First judge whether the text is a complete, self-contained utterance. Reject
isolated titles, section headings, pure numbers/codes, single-word fragments,
truncated quotations, or phrases that require missing context. Then judge
whether the text, without relying on vocal tone or context, does not clearly
express any particular emotion. Return JSON with
\texttt{semantic\_complete}, \texttt{is\_neutral}, \texttt{score}, and
\texttt{reason}; a higher score means more emotionally neutral.
\end{quote}
The challenging subset uses the salient-text prompt:
\begin{quote}\small
First judge whether the text is a complete, self-contained utterance. Then
judge whether the text, without relying on vocal tone, context, or speaker
background, clearly expresses the target emotion through the text itself.
Return JSON with \texttt{semantic\_complete}, \texttt{is\_salient},
\texttt{score}, and \texttt{reason}.
\end{quote}
For both modes, an utterance is selected only when the boolean decision is true
and the score is at least 70.

\paragraph{Style audio annotation.}
The style annotation prompt instructs the audio judge to evaluate delivery
style, not emotion or textual topic. It explicitly separates storytelling from
public-broadcast delivery, conversational delivery from performed delivery, and
intimate delivery from mere low volume. The judge assigns independent integer
scores from 0 to 4 for public-broadcast, intimate, dramatic, restrained-flat,
storytelling, and conversational, and returns \texttt{style\_scores},
\texttt{confidence}, and \texttt{rationale}. An utterance is selected when the
primary style score is at least 3, the primary--secondary margin is at least 1,
and the confidence exceeds the style-specific threshold: 0.70 for
public-broadcast, intimate, dramatic, and restrained-flat; 0.72 for
storytelling; and 0.78 for conversational.

\paragraph{Paralinguistic event annotation.}
The paralinguistic annotation prompt asks an audio judge to detect
speaker-produced non-verbal or semi-verbal events. It scores breath, laugh,
cough, and sigh independently on a 0--3 scale: 0 means absent, 1 faint, 2
noticeable, and 3 prominent. The judge returns \texttt{event\_scores},
\texttt{confidence}, and \texttt{rationale}. Add operations require the target
event to be absent or faint in the source utterance, while remove operations
require it to be noticeable or prominent.

\section{Compositional Task Construction}

Compositional samples are derived from verified atomic samples. One base
sample supplies the source audio and transcript, while other components supply
target attributes such as a speaker reference, an emotion label, a prosody
direction, or an acoustic environment. This makes each compositional instruction
act on a single source utterance while preserving atomic anchors for
component-level evaluation.

The compositional split uses four high-level attribute groups: semantic content,
speaker identity, expressive delivery, and acoustic environment. The split
contains the combinations in Table~\ref{tab:appendix-compositional-combos}.

\begin{table}[t]
\centering
\scriptsize
\setlength{\tabcolsep}{2pt}
\begin{tabular}{p{0.22\columnwidth}p{0.42\columnwidth}p{0.11\columnwidth}p{0.15\columnwidth}}
\toprule
Type & Combination & Count & EN/ZH \\
\midrule
2-component & \texttt{content+speaker} & 40 & 20/20 \\
2-component & \texttt{content+emotion} & 40 & 20/20 \\
2-component & \texttt{content+prosody} & 40 & 20/20 \\
2-component & \texttt{content+acoustic} & 40 & 20/20 \\
2-component & \texttt{speaker+emotion} & 40 & 20/20 \\
2-component & \texttt{speaker+acoustic} & 40 & 20/20 \\
2-component & \texttt{emotion+acoustic} & 40 & 20/20 \\
2-component & \texttt{prosody+acoustic} & 40 & 20/20 \\
\midrule
3-component & \shortstack[l]{\texttt{content+speaker+}\\\texttt{emotion}} & 20 & 10/10 \\
3-component & \shortstack[l]{\texttt{content+speaker+}\\\texttt{acoustic}} & 20 & 10/10 \\
3-component & \shortstack[l]{\texttt{content+emotion+}\\\texttt{acoustic}} & 20 & 10/10 \\
3-component & \shortstack[l]{\texttt{speaker+emotion+}\\\texttt{acoustic}} & 20 & 10/10 \\
\bottomrule
\end{tabular}
\caption{Compositional editing combinations and counts.}
\label{tab:appendix-compositional-combos}
\end{table}

Across the 400 compositional samples, component counts are content 220,
speaker 180, emotion 180, prosody 80, and acoustic 220. Prosody components in
the compositional split are speed and pitch controls, with 20 examples for each
of faster, slower, higher, and lower. Acoustic components use environment
transfer targets: bathroom 36, small-room 40, large-hall 40, outdoor 34, crowd
36, and music 34.

\section{Anchor Semantics}

Anchor-based evaluation defines what must be verifiable, without requiring a
unique target waveform. A target anchor specifies the intended edit, while a
preservation anchor specifies what should remain unchanged. Table~\ref{tab:appendix-anchor-semantics}
lists the anchor semantics used by the evaluator.

\begin{table*}[t]
\centering
\small
\setlength{\tabcolsep}{4pt}
\begin{tabular}{p{0.12\linewidth}p{0.39\linewidth}p{0.39\linewidth}}
\toprule
Task & Target anchor & Preservation anchor \\
\midrule
Content & Target transcript plus edit span and edit type. & The target transcript itself is the required content after editing; exact match and WER/CER are diagnostic. \\
Emotion & Target emotion label in the task taxonomy. & The source transcript should remain unchanged. \\
Style & Target style label in the six-style taxonomy. & The source transcript should remain unchanged. \\
Prosody & Direction and subtype: duration direction for speed, F0 direction for pitch, target words for stress. & The source transcript should remain unchanged. \\
Paralinguistic & Target event and operation, add or remove. & The source transcript should remain unchanged. \\
Speaker & Target-speaker reference clip and target speaker identity. & The source transcript should remain unchanged; source-output speaker similarity is additionally reported as a diagnostic outside speaker conversion targets. \\
Acoustic & Enhancement degradation type, or environment type and subtype with RT60/SNR metadata where applicable. & The source transcript should remain unchanged. Speaker similarity and perceptual quality are diagnostic. \\
Compositional & A list of atomic components, each with its own target anchor. & If no content component is present, the source transcript is preserved. If a content component is present, its target transcript becomes the expected content. \\
\bottomrule
\end{tabular}
\caption{Anchor semantics for target and preservation checks.}
\label{tab:appendix-anchor-semantics}
\end{table*}

A speaker reference is therefore not a waveform target; it is an identity
anchor. Similarly, an acoustic reference or acoustic condition describes a
measurable environment rather than a unique waveform. In compositional editing,
the components are evaluated independently and then combined into all-component
and joint success.

\section{Task Examples}

Table~\ref{tab:appendix-task-examples} shows representative task instances. The
examples report only task-relevant fields.

\begin{table*}[t]
\centering
\small
\setlength{\tabcolsep}{4pt}
\begin{tabular}{p{0.13\linewidth}p{0.31\linewidth}p{0.25\linewidth}p{0.22\linewidth}}
\toprule
Task & Source utterance excerpt & Instruction & Target anchor \\
\midrule
Content & ``... his belly counselled him.'' & Replace ``counselled'' with ``advised''. & Target transcript contains ``advised'' and no longer contains ``counselled''. \\
Emotion & ``The head of the Patchwork Girl was the most curious part of her.'' & Change the emotion to angry. & Target emotion is angry under the English emotion taxonomy. \\
Style & ``The singer being a man whose gifts lay chiefly in his throat and feet ...'' & Change the speaking style from storytelling style to public broadcast style. & Target style is public-broadcast. \\
Speaker & ``Perhaps it's because I haven't finished the first two races.'' & Convert the voice of the source audio to match the speaker in the reference clip. & Output should match the reference speaker identity. \\
Prosody & ``Once held by Hobson and Dewey, now carried by Mother Eddy and Brother Dowie.'' & Speak faster. & Output duration ratio must decrease by at least 5\%. \\
Paralinguistic & ``He had sinned mortally not once but many times ...'' & Add audible breathing sounds to the speech. & Breath score must reach the event-presence threshold. \\
Acoustic & ``Does a tiny particle of the consecrated bread contain all the body and blood ...'' & Remove background noise from the speech. & DNSMOS OVRL and BAK scores should improve over the degraded source. \\
Compositional, two-component & ``After early nightfall the yellow lamps would light up ...'' & Delete ``early'' and convert the voice to match the speaker in the reference clip. & Content deletion and target-speaker match must both succeed. \\
Compositional, three-component & ``An instant of wild flight had delivered him ...'' & Replace ``delivered'' with ``rescued'', convert the voice to match the speaker in the reference clip, and make the emotion sound frustrated. & Content replacement, target-speaker match, and target emotion must all succeed. \\
\bottomrule
\end{tabular}
\caption{Representative task examples.}
\label{tab:appendix-task-examples}
\end{table*}

\section{Evaluation Protocol Details}
\label{app:evaluation-details}

\subsection{Success Metrics}

For an evaluated sample set $\mathcal{D}$, let $t_i \in \{0,1\}$ denote target
success and $p_i \in \{0,1\}$ denote preservation success for sample $i$. Joint
success is the logical conjunction:
\[
j_i = t_i p_i .
\]
We abbreviate Target Success, Preservation Success, and Joint Success as
TS, PS, and JS:
\[
\mathrm{TS}=\frac{1}{|\mathcal{D}|}\sum_i t_i,
\]
\[
\mathrm{PS}=\frac{1}{|\mathcal{D}|}\sum_i p_i,
\]
\[
\mathrm{JS}=\frac{1}{|\mathcal{D}|}\sum_i j_i.
\]
For content editing, the edited transcript is itself the target, so the main
reported success is edit success; exact match and WER/CER are reported as
diagnostics rather than as an additional preservation gate.

\subsection{Content Preservation Gate}

For non-content atomic tasks, content preservation is a hard gate. The evaluator
first transcribes the output. English outputs use Whisper large-v3~\cite{radford2023robust,openai2023whisperlargev3}; Chinese
outputs use Paraformer~\cite{gao2022paraformer} through FunASR~\cite{gao2023funasr}. Text is normalized before comparison:
English is lowercased with punctuation removed, while Chinese removes spaces and
punctuation and preserves CJK characters, numbers, and Latin letters.

Let $y_i$ be the expected transcript and $\hat{y}_i$ be the ASR transcript. For
English, $e_i=\mathrm{WER}(y_i,\hat{y}_i)$; for Chinese,
$e_i=\mathrm{CER}(y_i,\hat{y}_i)$. The preservation gate is
\[
p_i = \mathbf{1}[e_i \leq 0.10].
\]
For non-content tasks, $y_i$ is the source transcript. For compositional samples
with a content component, $y_i$ is the content component's target transcript for
the content component itself; otherwise the source transcript is used for
preservation.

\subsection{Atomic Evaluators}

Table~\ref{tab:appendix-evaluator-thresholds} lists the target-success
criterion for each task. Diagnostic metrics are not used as substitutes for the
primary success criteria.

\begin{table*}[t]
\centering
\small
\setlength{\tabcolsep}{3.5pt}
\begin{tabular}{p{0.12\linewidth}p{0.46\linewidth}p{0.32\linewidth}}
\toprule
Task & Target-success criterion & Preservation and diagnostics \\
\midrule
Content & Replace succeeds if the target span is present and the original span is absent. Insert succeeds if the inserted span is present and, when an insertion anchor exists, appears after that anchor. Delete succeeds if the deleted span is absent. & Exact match, WER, and CER are reported. \\
Emotion & A Gemini-based multimodal judge predicts the dominant emotion from audio delivery only. Success requires the normalized predicted label to equal the target label; confidence requested in the judge response is not used as a threshold. & Content preservation gate with threshold 0.10. UTMOS and source-output speaker similarity are diagnostic. \\
Style & A target-conditioned Gemini-based multimodal judge scores the target style from 0 to 4. Success requires \texttt{target\_style\_success=true} and \texttt{target\_style\_score}$\geq 3$. If the boolean field is absent, score $\geq 3$ is used. & Content preservation gate with threshold 0.10. Six-dimensional style scores, UTMOS, and speaker similarity are diagnostic. \\
Prosody & Speed: faster requires output/source duration ratio $\leq 0.95$, slower requires $\geq 1.05$. Pitch: median F0 shift must be at least +0.3 semitone for higher and at most -0.3 for lower. Stress: all target words must be found by timestamp ASR and show positive prominence gain. & Content preservation gate with threshold 0.10. Continuous duration, F0, and prominence values are diagnostic. \\
Paralinguistic & A Gemini-based event judge scores each event from 0 to 3. Add succeeds if the target event score is at least 2. Remove succeeds if the target event score is at most 1. & Content preservation gate with threshold 0.10. Full event-score vectors are diagnostic. \\
Speaker & WavLM-large plus ECAPA-TDNN speaker embeddings are compared by cosine similarity. Success requires output-reference similarity $\geq 0.50$. & Content preservation gate with threshold 0.10. UTMOS is diagnostic. \\
Acoustic & Enhancement succeeds if DNSMOS OVRL gain $>0$ and DNSMOS BAK gain $>0$ over the degraded source. Reverb transfer succeeds if estimated RT60 falls within the target range with 0.8--1.2 tolerance. Noise transfer succeeds if the grouped PANNs scene prediction equals the target subtype, target score $\geq 0.10$, and target-score gain $\geq 0.03$. & Content preservation gate with threshold 0.10 for joint success. PESQ/STOI, UTMOS, and speaker similarity are diagnostic. \\
\bottomrule
\end{tabular}
\caption{Atomic evaluator target criteria and preservation gates.}
\label{tab:appendix-evaluator-thresholds}
\end{table*}

\subsection{Multimodal Judge Prompts}

The released emotion, style, and paralinguistic target evaluators are configured to use Gemini 2.5 Pro
(model identifier \texttt{gemini-2.5-pro}) through the Google
\texttt{v1beta/models/...:generateContent} endpoint. Each request contains the
task prompt followed by one inline, base64-encoded output-audio part with its
inferred MIME type. Decoding uses temperature 0 and requests a structured JSON
response (\texttt{application/json}). Requests have a 120-second timeout and up
to three attempts with exponential backoff. Responses are cached by caller,
model identifier, resolved audio path, and prompt; the parser accepts either raw
or fenced JSON. These settings improve repeatability, but the underlying
proprietary service may still change over time.

\paragraph{Emotion judge.}
The English prompt core is:
\begin{quote}\small
You are an expert speech-emotion evaluator. Judge dominant emotion from audio
delivery only. Do not infer from text semantics alone. Transcript: [ASR
transcript]. Allowed labels: [taxonomy labels]. Return JSON only with
\texttt{predicted\_emotion}, \texttt{confidence}, and \texttt{rationale}.
\end{quote}
The predicted label is normalized by alias mapping, for example fear to
fearful and surprised to surprise, and then matched against the target emotion.
The natural-language editing instruction is not supplied to this judge; it
receives the output audio, its ASR transcript, and the allowed language-specific
emotion taxonomy.

\paragraph{Style judge.}
The target-conditioned style prompt asks the judge to listen to the edited audio
and decide whether the target speaking style is clearly present. It states:
\begin{quote}\small
Evaluate only vocal delivery: projection, prosody shape, rhythm, pacing, and
speaking manner. Do not judge from the text content, topic, or word meanings. Do
not annotate emotion categories. Even if the transcript sounds like a story,
news, or dialogue, base the judgment on audio delivery.
\end{quote}
The JSON output includes \texttt{target\_style\_score},
\texttt{target\_style\_success}, \texttt{dominant\_style},
\texttt{style\_scores}, \texttt{confidence}, and \texttt{rationale}.
The prompt includes the target style and its definition, making this evaluator
target-conditioned rather than an unconstrained style classifier.

\paragraph{Paralinguistic judge.}
The event judge listens for breath, laugh, cough, and sigh. It is instructed to
score only speaker-produced vocal events, not background noise or microphone
artifacts. The JSON output contains a four-dimensional \texttt{event\_scores}
object, \texttt{confidence}, and \texttt{rationale}. The same score vector is
used for both add and remove decisions.

\subsection{Human Validation of Judge-Dependent Metrics}
\label{app:human-validation}

\paragraph{Sampling.}
We sampled 240 outputs: 80 each from emotion, style, and paralinguistic editing.
Sampling used seed 42 and stratified round-robin selection over system, automatic
target-success status, language, and task subtype. The subset spans nine systems
whose output audio was available; GLM-4-Voice appears only in this validation
analysis and is not included in the eight-system Speech LLM comparison in
Table~\ref{tab:main-results}. The 27 task--system cells contain 3--14 outputs
each, so cell-level correlations are informative but should not be interpreted
as estimates from a perfectly balanced factorial sample.

\paragraph{Annotation protocol.}
Ten annotators each labeled 72 items, for three independent annotations per output
and 720 item-level annotations in total. Assignment bundles concealed system identity,
automatic decisions, sampling strata, and internal identifiers. Annotators were
shown the source and edited audio, task, target, language, instruction, and
source transcript. They answered two binary questions: whether the target edit
succeeded and whether the edited output preserved the linguistic content. The
content question explicitly excluded speaker timbre. Human joint success is the
conjunction of these two answers. We use majority vote for automatic--human
comparisons, mean pairwise agreement and Krippendorff's $\alpha$ for
human--human reliability, and Pearson/Spearman correlations over task--system
cell rates.

Tables~\ref{tab:human-validation-overall} and
\ref{tab:human-validation-task} report overall and task-level statistics,
respectively. Table~\ref{tab:human-validation-cases} shows representative
agreements and disagreements.

\begin{table*}[t]
\centering
\small
\setlength{\tabcolsep}{4.2pt}
\begin{tabular}{llcccccc}
\toprule
Scope & Criterion & Auto--human & Balanced & F1 & $\kappa$ & Pearson $r$ & Spearman $\rho$ \\
& & agreement (\%) & acc. (\%) & (\%) & & & \\
\midrule
Overall & Target & 60.4 & 60.5 & 59.9 & .209 & .270 & .323 \\
Overall & Joint & 68.8 & 63.9 & 48.3 & .302 & .685 & .699 \\
\bottomrule
\end{tabular}
\caption{Overall automatic--human agreement against majority labels. Correlations are computed over 27 task--system cells. For joint success, Pearson $p=8.0\times10^{-5}$ and Spearman $p=4.9\times10^{-5}$.}
\label{tab:human-validation-overall}
\end{table*}

\begin{table*}[t]
\centering
\small
\setlength{\tabcolsep}{3.8pt}
\begin{tabular}{lccccccc}
\toprule
Task & Target auto--human & Joint auto--human & Joint $\kappa$ & Joint $r$ & Joint $\rho$ & Joint human--human & Joint $\alpha$ \\
& agree. (\%) & agree. (\%) & & & & agree. (\%) & \\
\midrule
Emotion & 51.3 & 65.0 & .231 & .525 & .567 & 75.8 & .504 \\
Style & 62.5 & 67.5 & .342 & .743 & .778 & 81.7 & .633 \\
Paralinguistic & 67.5 & 73.8 & .291 & .731 & .676 & 83.3 & .636 \\
\midrule
Overall & 60.4 & 68.8 & .302 & .685 & .699 & 80.3 & .595 \\
\bottomrule
\end{tabular}
\caption{Task-level validation statistics ($n=80$ per task). Human--human values are mean pairwise agreement and Krippendorff's $\alpha$ over the joint labels.}
\label{tab:human-validation-task}
\end{table*}

\begin{table*}[t]
\centering
\small
\setlength{\tabcolsep}{4pt}
\begin{tabular}{p{0.13\linewidth}p{0.15\linewidth}p{0.28\linewidth}p{0.15\linewidth}p{0.15\linewidth}}
\toprule
Case & System & Task and instruction & Automatic T/C/J & Human-majority T/C/J \\
\midrule
Agreement: success & Gemini-Live & Emotion (EN): change to happy & 1/1/1 & 1/1/1 \\
Agreement: failure & GPT-Realtime & Emotion (EN): change to frustrated & 0/0/0 & 0/0/0 \\
Automatic false positive & Kimi-Audio & Emotion (ZH): change to fear & 1/1/1 & 0/1/0 \\
Automatic false negative & Gemini-Live & Emotion (EN): change to surprise & 0/1/0 & 1/1/1 \\
\bottomrule
\end{tabular}
\caption{Representative validation cases. T/C/J denote target success, linguistic-content preservation, and their conjunction. The cases illustrate agreement as well as both directions of judge error without attributing disagreement to an unverified perceptual cause.}
\label{tab:human-validation-cases}
\end{table*}

\subsection{Diagnostic Metrics}

UTMOS is reported as a naturalness diagnostic using the UTMOS22 quick-prediction
model~\cite{saeki2022utmos}. Speaker similarity is computed by extracting WavLM-large hidden states \cite{chen2022wavlm},
feeding them to an ECAPA-TDNN speaker head~\cite{dawalatabad2021ecapa}, and taking cosine similarity between
embeddings. DNSMOS follows the Microsoft DNSMOS P.835 model~\cite{reddy2022dnsmos} and reports
SIG, BAK, and OVRL. PESQ and STOI are computed only when a
clean reference is available and are treated as diagnostics. For acoustic
environment transfer, PESQ/STOI are not primary success metrics because a
successful environment transfer may intentionally reduce clean-speech quality
scores.

\section{Compositional Metrics}

Let $C_i$ be the set of components in compositional sample $i$, and let
$a_{i,c}\in\{0,1\}$ indicate whether component $c$ succeeds under the
corresponding atomic evaluator. The pooled component success (CS) is
\[
\mathrm{CS} =
\frac{\sum_i \sum_{c\in C_i} a_{i,c}}
     {\sum_i |C_i|}.
\]
The all-component indicator is
\[
q_i = \prod_{c\in C_i} a_{i,c}.
\]
The compositional joint indicator is
\[
j_i = q_i p_i,
\]
where $p_i$ is the preservation result. Thus component success measures
individual sub-goal achievement, all-component success measures whether all
requested edits are satisfied simultaneously, and joint success further requires
preservation of non-target content when applicable.

\section{Failure-Type Taxonomy}

Table~\ref{tab:appendix-failure-types} summarizes representative failure types
observed during evaluation. The examples are diagnostic descriptions rather
than additional headline results.

\begin{table*}[t]
\centering
\small
\setlength{\tabcolsep}{3.5pt}
\begin{tabular}{p{0.15\linewidth}p{0.28\linewidth}p{0.30\linewidth}p{0.17\linewidth}}
\toprule
Failure type & Example instruction & Observed metric pattern & Interpretation \\
\midrule
Target not applied & Change the emotion to angry. & Judge predicts neutral; preservation error is 0.000. & The model preserves content but does not realize the requested expressive target. \\
Wrong expressive label & Change from dramatic style to public broadcast style. & Dominant style remains dramatic; content preservation passes. & The output keeps the source delivery style rather than moving to the target style. \\
Local content edit with global drift & Replace a named phrase in a Chinese utterance. & The edit span can be present, but sentence-level CER is extremely high. & Local lexical control can coexist with severe unintended transcript drift. \\
Speaker target miss & Convert the voice to match the reference clip. & Output-reference similarity remains below 0.50, often with large ASR error. & The model either ignores the reference identity or changes content while attempting conversion. \\
Prosody too weak or wrong direction & Raise the pitch; speak faster; emphasize a target word. & F0 shift is below 0.3 semitone, duration ratio does not cross the 5\% threshold, or target-word prominence does not increase. & The requested prosodic control is absent, too small, or not localized. \\
Paralinguistic add/remove miss & Add audible breathing; remove audible breathing. & Add has event score 0 or 1; remove has event score 2 or 3. & The target event is either not introduced or not removed. \\
Acoustic target without preservation & Remove reverberation. & DNSMOS gain can be positive while content-preservation error is far above 0.10. & Enhancement-like processing improves acoustic scores but damages intelligibility. \\
Environment target miss & Add background music or crowd noise. & PANNs target scene score is below 0.10 or target gain is below 0.03. & The output does not contain a measurable target environment. \\
Compositional partial success & Delete a word and convert speaker identity. & Content component succeeds, but speaker similarity fails; all-component and joint success are false. & Multi-objective instructions often fail by satisfying only one component. \\
\bottomrule
\end{tabular}
\caption{Representative failure-type taxonomy used for qualitative analysis.}
\label{tab:appendix-failure-types}
\end{table*}

\section{AI Usage}

For manuscript preparation, we used AI-assisted tools for grammatical checking.

\section{Model Size}

Ming-UniAudio-16B-A3B-Edit has 16 billion parameters, Step-Audio-EditX has 3 billion, and Qwen3-Omni has 30 billion. Kimi-Audio, Mimo-Audio-Base, and Mimo-Audio-Instruction each have 7 billion parameters. Seed-VC has 98 million parameters, and VoxCPM2 has 2 billion. Local generation and evaluation use CUDA-enabled NVIDIA GPUs; hardware varies across model-specific environments. We use each released model interface with the task-specific settings recorded by the corresponding runner and generation metadata.

For specialized content references, the reported English and Chinese VoiceCraft-X results use the English checkpoint and the Chinese \texttt{voicecraftx-zh.ckpt} checkpoint with Chinese text normalization, respectively. Chatterbox event addition uses native Turbo tags for English; Chinese addition uses the multilingual model with inline tags as a best-effort configuration. AudioSep supplies the event-removal outputs. These configurations are task-specific reference pipelines rather than unified source-speech editors.

\section{Licenses}

The benchmark uses artifacts with both open licenses and restricted access terms. LibriTTS, WenetSpeech, VCTK, and MUSAN are distributed under CC BY 4.0; LibriQuote uses CC BY-NC 4.0. AISHELL-3, NonverbalTTS, CSEMOTIONS, and RIRS\_NOISES use Apache 2.0, while AISHELL6-Whisper and Emilia-NV use CC BY-NC-SA 4.0. MagicData-RAMC uses a custom open-data license, IEMOCAP uses a custom access agreement, and StoryTTS requires its custom research-use agreement. The DisfluencySpeech and NaturalVoices repositories license their code under MIT; NaturalVoices separately records source-dependent audio licenses, which apply to the individual audio files.

For evaluated open models and code, Ming-UniAudio and Chatterbox use MIT; Step-Audio-EditX, Qwen3-Omni, MiMo-Audio, and VoxCPM2 use Apache 2.0; Seed-VC uses GPL-3.0; DeepFilterNet is dual-licensed under MIT and Apache 2.0; and VoiceCraft-X uses CC BY-NC 4.0. Kimi-Audio uses mixed licensing: Qwen-derived code is Apache 2.0 and the remaining released code is MIT. Model weights and third-party components remain subject to their accompanying terms. Gemini-Live and GPT-Realtime are proprietary API services governed by platform terms.